\documentclass[aps,prc,twocolumn,amsmath,amssymb,superscriptaddress,showpacs,showkeys,preprintnumbers,nofootinbib]{revtex4-2}
\usepackage{dcolumn}
\usepackage{graphicx,color}
\usepackage{multirow}
\usepackage{textcomp}
\usepackage{physics}
\usepackage{dsfont}
\usepackage{relsize}

\usepackage{tikz}	
\begin{document}
	\newcommand {\nc} {\newcommand}
	\nc {\beq} {\begin{eqnarray}}
	\nc {\eeq} {\nonumber \end{eqnarray}}
	\nc {\eeqn}[1] {\label {#1} \end{eqnarray}}
\nc {\eol} {\nonumber \\}
\nc {\eoln}[1] {\label {#1} \\}
\nc {\ve} [1] {\mbox{\boldmath $#1$}}
\nc {\ves} [1] {\mbox{\boldmath ${\scriptstyle #1}$}}
\nc {\mrm} [1] {\mathrm{#1}}
\nc {\half} {\mbox{$\frac{1}{2}$}}
\nc {\thal} {\mbox{$\frac{3}{2}$}}
\nc {\fial} {\mbox{$\frac{5}{2}$}}
\nc {\la} {\mbox{$\langle$}}
\nc {\ra} {\mbox{$\rangle$}}
\nc {\etal} {\emph{et al.}}
\nc {\eq} [1] {(\ref{#1})}
\nc {\Eq} [1] {Eq.~(\ref{#1})}
\nc {\Refc} [2] {Refs.~\cite[#1]{#2}}
\nc {\Sec} [1] {Sec.~\ref{#1}}
\nc {\chap} [1] {Chapter~\ref{#1}}
\nc {\anx} [1] {Appendix~\ref{#1}}
\nc {\tbl} [1] {Table~\ref{#1}}
\nc {\Fig} [1] {Fig.~\ref{#1}}
\nc {\ex} [1] {$^{#1}$}
\nc {\Sch} {Schr\"odinger }
\nc {\flim} [2] {\mathop{\longrightarrow}\limits_{{#1}\rightarrow{#2}}}
\nc {\textdegr}{$^{\circ}$}
\nc {\inred} [1]{\textcolor{red}{#1}}
\nc {\inblue} [1]{\textcolor{blue}{#1}}
\nc {\IR} [1]{\textcolor{red}{#1}}
\nc {\IB} [1]{\textcolor{blue}{#1}}
\nc{\pderiv}[2]{\cfrac{\partial #1}{\partial #2}}
\nc{\deriv}[2]{\cfrac{d#1}{d#2}}

\nc {\bit} {\begin{itemize}}
	\nc {\eit} {\end{itemize}}

\title{Quantifying uncertainties due to optical potentials in one-neutron knockout reactions}
\author{C.~Hebborn}
\email{hebborn@frib.msu.edu}
\affiliation{Facility for Rare Isotope Beams, Michigan State University, East Lansing, Michigan 48824, USA}
\affiliation{Lawrence Livermore National Laboratory, P.O. Box 808, L-414, Livermore, California 94551, USA}
\author{T. R. Whitehead}
\affiliation{Facility for Rare Isotope Beams, Michigan State University, East Lansing, Michigan 48824, USA}
\author{A.~E.~Lovell}
\affiliation{Theoretical Division, Los Alamos National Laboratory, Los Alamos, New Mexico 87545, USA}
\author{F.~M.~Nunes}
\affiliation{Facility for Rare Isotope Beams, Michigan State University, East Lansing, Michigan 48824, USA}
\affiliation{Department of Physics and Astronomy, Michigan State University, East Lansing, Michigan 48824, USA}

\date{\today}
\preprint{LLNL-JRNL-842695, LA-UR-22-32049}
\begin{abstract}
One-neutron knockout reactions have been widely used to extract information about the single-particle structure of nuclei from the valley of stability to the driplines.
The interpretation of knockout data relies on reaction models, where the uncertainties are typically not accounted for. In this work we quantify uncertainties of optical potentials used in these reaction models and propagate them, for the first time, to knockout observables using a Bayesian analysis. We study  two reactions in the present paper, the first of which involves a  loosely-bound halo projectile, $^{11}$Be, and the second a tightly-bound projectile, $^{12}$C. 
We first quantify  the parametric uncertainties associated with phenomenological optical potentials. Complementing to this approach, we also quantify the model uncertainties associated with the chiral forces that can be used to construct microscopic optical potentials. 
For the phenomenological study, we investigate the impact of the imaginary terms of the optical potential on the breakup and stripping components of the knockout cross sections as well as the impact of the angular range. For the $^{11}$Be case, the theoretical uncertainty from the phenomenological method is on the order of the experiment uncertainty on the knockout observables; however, for the $^{12}$C case, the theoretical uncertainty is significantly larger. The widths of the uncertainty bands for the knockout observables obtained for the microscopic study and the phenomenological approach are of similar order of magnitude. Based on this work we conclude that structure information inferred from the ratio of the knockout cross sections, will carry a theoretical uncertainty of at least $20\%$ for halo nuclei and at least $40\%$ for tightly-bound nuclei. 
\end{abstract}

\maketitle
%

\section{Introduction}

It is an extraordinary time for rare isotope science, with new generation isotope facilities coming on-line and enabling laboratory exploration of a wider region in the nuclear chart, particularly along the proton dripline and much closer to the neutron dripline than before. Reaction studies are a preferred tool to explore these most exotic species, with transfer, breakup and knockout being amongst the most commonly used. While both transfer and breakup measurements require significant beam intensities, knockout experiments can be performed with $\approx 100$ pps beams because they entail an inclusive measurement. In single-nucleon knockout experiments, the core fragment is detected in coincidence with $\gamma$ rays following its de-excitation, allowing for the isolation of the different  single-particle components in the original projectile state. Information regarding the angular momentum is extracted from the momentum distributions of the core fragment, while spectroscopic factors (SF) for each component are obtained from the comparison of the total measured cross section and the corresponding theoretical prediction. 

Over the last two decades, experimentalists have collected a large body of knockout data, spanning a wide range of proton-neutron asymmetry (see \cite{PhysRevC.90.057602,PhysRevC.103.054610} for a recent compilation). Disagreements between the extracted SFs and those predicted by large scale shell-model calculations were found for many of the systems considered. A systematic analysis of the ratio of the SFs extracted from knockout measurements and those predicted by theory revealed a strong dependence on the neutron to proton asymmetry in the system. Indeed, a linear dependence on the difference between the neutron and proton separation energies in the system was found for that SF ratio. The same quantity extracted using other probes, such as quasi-free scattering ~\cite{GM18} and transfer reactions~\cite{Tetal09,LeePRC06}, do not show the same trend. Although several groups have studied thoroughly the validity of the theoretical models involved for the last two decades,
the origin of this discrepancy is still unsettled and needs to be resolved (see recent review in Ref.~\cite{AUMANN2021103847}).

It is clear that the reliability of the nuclear-structure information extracted  from any reaction probe depends on the reaction model used. For knockout experiments, the eikonal model, the standard method used \cite{G59,HM85,HT03}, has been studied in some detail (e.g. \cite{PhysRevC.66.024607}). 
Like other reaction models, the inputs to the eikonal model, namely the fragment-target interactions, carry uncertainties.
The goal of the current study is to quantify the optical potential uncertainties in predictions for knockout observables, to better inform their  interpretation.

 Significant work has been undertaken to quantify, using Bayesian analyses, the uncertainties coming from the fit of the potential parameters to experimental scattering data  (e.g. \cite{PhysRevLett.122.232502,Lovelletal21}). The propagation of these uncertainties  to transfer observables \cite{PhysRevLett.122.232502,Lovelletal21} and charge-exchange observables \cite{PhysRevC.105.054611} shows that for some cases the uncertainties on the cross section angular distributions are much larger than the previous rough estimates of~$30$\%. In the present article, we extend those studies to determine how the uncertainties in the optical potential propagate to knockout observables. We focus on uncertainties from the nucleon-target and core-target optical potentials, and for now do not consider uncertainties in the bound-state description (recent studies explored the sensitivity of knockout observables to the bound state effective interaction \cite{PhysRevC.100.054607,HCinprep}).

Complementing the phenomenological approach of \cite{PhysRevLett.122.232502,Lovelletal21}, in the current study we also quantify uncertainties in knockout when using microscopic optical potentials derived from many-body calculations. In the last decade, theorists have made important progress in extracting \textit{ab initio} optical potentials in which the crucial element is the nuclear force~(see Ref. \cite{whitepaper} for a recent review). In particular, the nucleon optical potential was derived from nuclear matter calculations considering a set of five different nuclear forces derived within chiral effective field theory (EFT) \cite{Whitehead21}. Recently, a formulation for the nucleus-nucleus optical potential  derived from the same chiral force and the energy densities
of the interacting nuclei was implemented \cite{whitehead2022,Brueckner68}. We combine these two microscopic developments and include in this paper a study of theoretical uncertainties on knockout observables coming directly from the underlying chiral force through the microscopic optical model derived in nuclear matter.

This paper is organized in the following manner. In Sec.~\ref{Sec2}, we briefly discuss the methodology and provide numerical details. In Sec.~\ref{Sec3}, results for the phenomenological approach are presented and discussed. Sec.~\ref{Sec4} contains the results following the microscopic approach. Finally, in Sec.~\ref{Sec5}, conclusions are drawn.

\section{Methodology and numerical details }\label{Sec2}
\subsection{Reaction model}
In this article, we consider one-neutron knockout reactions on a $^9\rm Be$ target at 60$A$~MeV, which are the typical target and beam energy used in experiments. As usual, we evaluate the knockout cross section in a few-body framework, in which the projectile is seen as a structureless core $c$ and a valence neutron $n$ impinging on a structureless target $T$~\cite{HT03,BC12}. The inputs of this few-body model are the single-particle overlap function and effective core- and neutron-target interactions, so-called optical potentials, accounting for all non-elastic channels not explicitly included in the theory. Apart from the optical potentials, knockout observables are then sensitive mainly to the projectile separation energy $S_n$, the spin, parity and root mean square (rms) radius $\sqrt{\langle r^2\rangle}$ of the projectile overlap function~\cite{HT03,PhysRevC.90.057602,PhysRevLett.119.262501,HCinprep}.

 To  quantify the uncertainties due to the choice of optical potentials in knockout reactions, we consider two different projectiles which have different $S_n$ and $\langle r^2\rangle$. First, we study the reaction involving a halo nucleus projectile, $^{11}\rm Be$, the ground 1/2$^+$ and 1/2$^-$  excited states of which are described as a $^{10}$Be inert core to which a valence neutron is bound  in a $1s_{1/2}$ orbital by $S_n=500$~keV and in a $0p_{1/2}$ orbital by 184 keV, respectively~\cite{KELLEY201288}. Second, we consider a $^{12}$C projectile, which in its ground state is seen as a $^{11}$C inert core with a valence neutron in a $0p_{3/2}$ bound by $S_n=$18.72 MeV. 
 For both cases, we neglect the spin of the core, and we use single-particle Gaussian potentials adjusted to reproduce the experimental $S_n$ and rms radius ($\sqrt{\langle r^2\rangle}$=6.5$\pm 0.3$ fm for $^{11}$Be~\cite{11Berms} and $\sqrt{\langle r^2\rangle}$=2.35-2.48~fm for $^{12}$C~\cite{KELLEY201771}). Note that as for halo nuclei, the asymptotic normalization constant (ANC) and the rms radius are strongly correlated, enforcing the experimental rms radius enables the reproduction of predicted~\cite{PhysRevLett.117.242501} and evaluated ANCs~\cite{PhysRevC.98.054602,PhysRevC.104.024616}.

We compute the two contributions to the knockout cross section $\sigma_{ko}$, i.e. the diffractive breakup $\sigma_{bu}$ and stripping $\sigma_{str}$, using the eikonal model{~\cite{PhysRevC.54.3043,HT03,HM85}. The total cross sections are obtained by integrating  the momentum distributions.} The Coulomb interaction is treated using the correction described in Refs.~\cite{MBB03,CBS08}. These two contributions correspond to different reaction processes, i.e., the diffractive breakup to the reaction channel in which both the core and the neutron survive the collision, while the stripping to the channel in which the neutron is absorbed by the target. The relative importance of $\sigma_{bu}$ increases for projectiles with smaller $S_n$, since the projectile is more fragile and breaks up more easily. 
For the $^{11}$Be ($^{12}$C) calculations, we use the following model space: the $^{10}$Be-$n$ ($^{11}$C-$n$) continuum is described up to the $c$-$n$ orbital angular momentum $l_{\rm max} = 12$ ($l_{\rm max} = 6$).
 
\subsection{Bayesian uncertainty quantification }~\label{Sec2B}

When effective interactions are employed in the few-body reaction framework, their inherent uncertainties propagate to the reaction observables of interest. In the case of phenomenological effective interactions, one source of uncertainty arises from optimizing the model parameters to experimental data. Optical potentials are usually expressed in the following form
\begin{align}
\label{Eq1}
U(r) &= -V f(r;R_V,a_V) - i W f(r;R_W,a_W) \\ \nonumber 
&+i 4a_S W_S \frac{d}{dr} f(r;R_S,a_S) \\ \nonumber
\end{align}
where $f(r;R_i,a_i)=\frac{1}{1+e^{(r-R_i)/a_i}}$ and $V$, $W$, $W_S$ are the real volume, imaginary volume, and imaginary surface depths, $R_i = r_i A_T^{1/3}$ and $R_i=r_i (A_c^{1/3}+ A_T^{1/3})$ respectively for the neutron- and core-target potential, where $A_c$ and $A_T$ denote the mass numbers of the core and the target. 

To quantify these parametric uncertainties we follow a Bayesian approach \cite{Lovell18,PhysRevC.100.064615,Whitehead22a} that assumes a prior distribution and then explores the parameter space guided by comparisons to experimental data. After a sufficiently long exploration of parameter space, a posterior distribution of the parameters is obtained. This posterior may be interpreted as the most likely parameter distributions of the effective interaction given the prior knowledge and the experimental data considered in the likelihood.

In the current work, we optimize the effective interactions to elastic scattering data and then calculate knockout cross sections with parameter sets sampled from the posterior distributions. 
Since there are no $^{10}$Be-$^{9}$Be, $^{11}$C-$^{9}$Be and $n$-$^9$Be scattering data available, we considered here mock data, that we generate at every 1$^\circ$ using realistic interactions (we have verified that we obtain similar posterior distribution for mock data generated at every $3^\circ$). For this, we use the optical potential that was fit to $^{10}$Be-$^{12}$C scattering at 59.4$A$ MeV \cite{PhysRevC.55.R1018} and a $n$-$^{9}$Be potential fitted to reproduce elastic-scattering and polarization observables for a nucleon off a target nucleus with $A_T\leq 13$ at energies between 65 MeV and 75 MeV~\cite{Weppner18}.
The parameters of both interactions used to generate the mock data are listed in Table~\ref{t1}~\footnote{The spin orbit part of $n$-$^9$Be potential was not considered in this study.}. We assign an error of 10\% to all mock data, which is common for elastic-scattering  experiments with stable beams. 
We assume prior Gaussian distributions centered on parameter values used to produce the mock data with widths equal to the mean values. This choice of a large prior width is made to allow for a data-driven posterior. Unreasonable parameters values, such as a $r_i < 0, a_i < 0, W < 0$, are excluded. 

For the Bayesian analysis, we follow Ref.~\cite{Lovell18} and we use a Metropolis-Hastings Markov chain Monte Carlo (MCMC). To ensure that initial parameters are within the posterior that we are interested in sampling, we begin the search with a burn-in phase, as described in Ref.~\cite{Lovell18} and using $N_{burn-in}=500$.
Once a good region of the parameter space is found, random steps, with size scaled to $\epsilon=0.02$ of the parameter prior means, are taken in parameter space. To guarantee independence of the parameter samples, every tenth sample is recorded. The resulting posterior, formed from 16 parallel parameter searches that each collect 1600 parameter sets, is used to build 68\% and 95\% uncertainty intervals for elastic-scattering cross sections. We then draw 1600 random samples from this posterior to generate the uncertainty bands for knockout cross sections.

\begin{table}
\begin{tabular}{ccccccccccc} \hline \hline
&$V$ & $r_V$& $a_V$ & 
 $W$ & $r_W$ &$a_W$ &$W_S$ & $r_S$ &$a_S$ &$r_C$ \\
 &[MeV]&[fm]&[fm]&[MeV] & [fm]&[fm]&[MeV]& [fm]&[fm]&[fm]\\ 
\multirow{1}{*}{$c$-$^{9}\mathrm{Be}$}&123.0& 0.75& 0.80& 65.0& 0.78& 0.80& &&&1.2 \\
\multirow{1}{*}{$n$-$^{9}\mathrm{Be}$}&33.08 & 1.14& 0.65&4.15& 1.14 & 0.65 &9.18 &1.36 &0.18& \\ \hline 		\hline
\end{tabular}	
\caption{Parameters of the optical potentials used to generate the $c$-$^9\rm Be$ and $n$-$^9\rm Be$ mock data considered in the Bayesian analysis for the one-neutron knockout of $^{11}\rm Be$ and $^{12}$C on $^9\rm Be$ at 60$A$~MeV. 	\label{t1}}
\end{table}

\section{Phenomenological approach} \label{Sec3}

In this section, we study the parametric uncertainties associated with the phenomenological potentials, arising from fitting to elastic scattering data. We apply a Bayesian analysis to all optical potential parameters (9 for the $n$-$^9$Be potential and 6 for the $c$-$^9$Be potential). As discussed in Appendix~\ref{AppA}, we observe that the geometry of the imaginary volume terms of both interactions are poorly constrained by elastic-scattering data, and that adding reaction cross section data did not improve their constraint. To avoid unconstrained parameters, we impose the imaginary volume geometry to be equal to the real volume geometry, i.e., $r_W=r_V$ and $a_W=a_V$, a simplification consistent with Ref.~\cite{Koning2003}. 

\subsection{Influence of the backwards scattering angles}
\label{Sec3A}

We first investigate the influence of $n$-$^9\rm Be$ scattering data at backwards scattering angles (a similar study for the $c$-$^9\rm Be$ interaction can be found in Appendix~\ref{AppB}). 
In Fig. \ref{Fig2}, we compare two Bayesian analyses of the $n$-$^9$Be interaction constrained with mock elastic-scattering data at forward angles $\in [0,60^\circ]$ (f., in blue) and including data at backwards angles $\in [0,120^\circ] $ (f.+b., in magenta). As expected, including data at backwards angles significantly narrows the 68\% (hatched area) and 95\% (dotted lines) uncertainty bands of elastic-scattering cross section at these angles [Fig. \ref{Fig2}(b)]. However, the impact of these data onto the parameter posteriors and their correlations is much more modest, as seen in the corner plot in Fig. \ref{Fig2}(a). Indeed, the additional data reduce only slightly the widths of all posteriors [diagonal of Fig. \ref{Fig2}(a)], indicating that they do not greatly improve the constraint of the optical potential parameters. Moreover, only the surface term parameters [off-diagonal plots on the bottom right part of Fig. \ref{Fig2}(a)] exhibit a significantly different correlation structure.

\begin{figure*}[hbt!]
	\centering

		{\includegraphics[width=\linewidth]{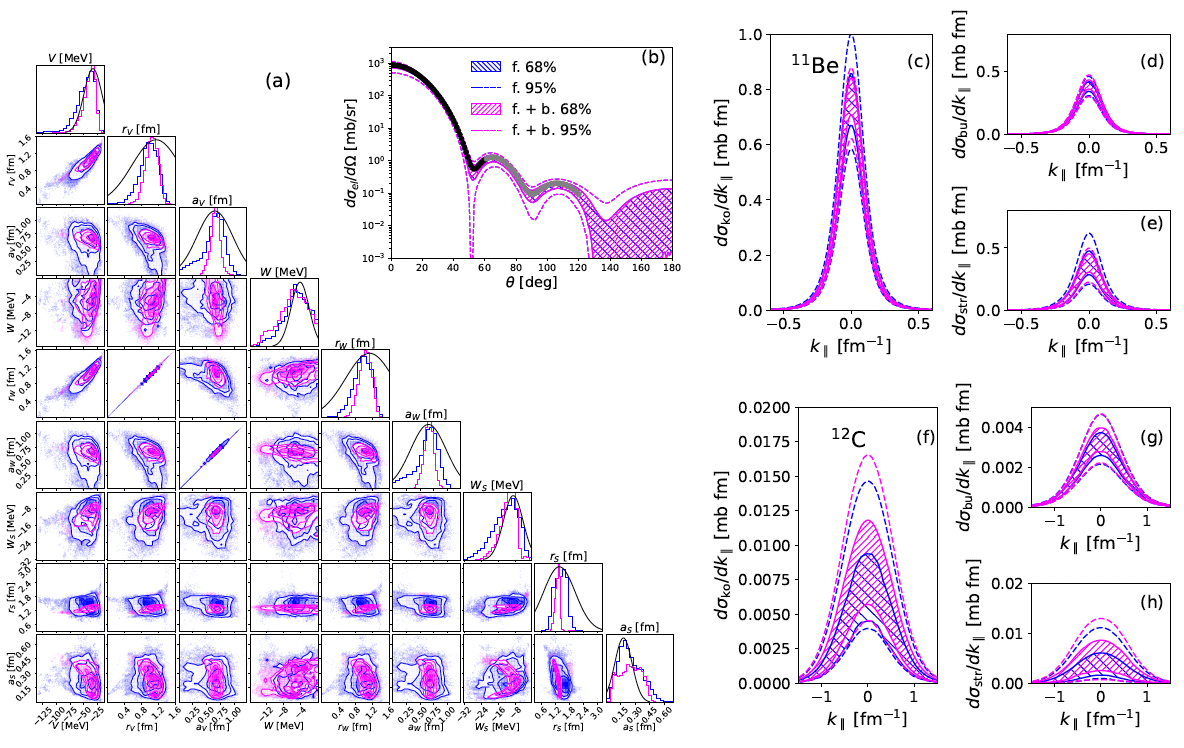}}

	\caption{Influence of the backwards scattering angles for the $n$-$^9\rm Be$ interaction onto elastic and knockout observables: the optical potential parameters are constrained with data $\in [0,60^\circ]$ (f., in blue) and $\in [0,120^\circ]$ (f. + b., in magenta). (a) Corner plots; (b) Elastic-scattering cross sections of $n$ off $^{9}\rm Be$ as a function of the scattering angle $\theta$ with mock data $\in [0,60^\circ]$ (black points) and $\in [60^\circ,120^\circ]$ (gray points); One-neutron knockout cross sections, (c) $\sigma_{ko}$ along with their (d) diffractive-breakup $\sigma_{bu}$ and (e) stripping $\sigma_{str}$ contributions, for $^{11}\rm Be$ and {(f,g,h) $^{12}$C off a $^9$Be target at $60A$~MeV.} }\label{Fig2}

\end{figure*}

 For the knockout of a loosely-bound neutron, i.e., the one-neutron knockout of $^{11}$Be, 
the stripping $\sigma_{str}$ [Fig. \ref{Fig2}(e)] and diffractive-breakup $\sigma_{bu}$ [Fig.~\ref{Fig2}(d)] contributions are of the same order. Interestingly, 
 including data at backwards angles reduces strongly the 95\% uncertainty band on stripping observables, but does not influence the diffractive-breakup observables. Consequently, the 95\% uncertainty interval on the total knockout cross sections [Fig. \ref{Fig2}(c)] is slightly narrowed. As the neutron is absorbed by the target in the stripping process, this suggests that  these additional data mainly affect the absorptive strength of the $n$-$^9$Be interaction. Although the 95\% uncertainty intervals are reduced, the 68\% uncertainty bands for all cross sections are not impacted by these data at backwards angles, indicating of the complex structure of the likelihood in this region of the parameter space.

 The knockout of a more deeply-bound neutron shown in Fig. \ref{Fig2}(f), i.e., the one-neutron knockout of $^{12}$C, is dominated by the stripping contribution [Fig. \ref{Fig2}(h)]. As for the loosely-bound case, including data at backwards scattering angles does not influence $\sigma_{bu}$ [Fig. \ref{Fig2}(g)]. However, contrary to the loosely-bound case, these additional data impact both the 68\% and 95\% intervals obtained for stripping observables, and hence for total knockout cross sections: both their magnitudes and their widths are larger when backwards angles are included. 
 
 These contrasting effects for the removal of loosely- and more deeply-bound neutrons can be understood by the mechanisms at play during the reaction, which are different for both contributions. 
 On one hand, $\sigma_{bu}$ is mostly influenced by forward scattering angles, because most of the breakup will occur at impact parameters of the neutron that are large enough to avoid its absorption by the target. Semiclassically, these large impact parameters can be interpreted as forward scattering angles. 
On the other hand, as the neutron is absorbed by the target in the stripping process, it takes place at small impact parameter, i.e. large $n$-$^9$Be scattering angles. Moreover, the effect on stripping observables depends strongly on the separation energy of the projectile, as the core must survive the collision. 
For deeply-bound neutrons at small impact parameters, the core is also close to the target and can be absorbed. The role of $n$-$^9$Be elastic data at backwards scattering angles in knockout cross sections is therefore intertwined with the range of the imaginary part of the core-target optical potential, which  drives the probability that the core survives the collision. As shown in Appendix~\ref{AppB}, we observe a similarly complicated influence of the backwards angles for the $c$-$^9$Be interaction in knockout observables.

Consequently, because knockout distributions have a complicated dependence on both the $n$- and $c$-$^9$Be optical potentials, realistic estimates of their parametric uncertainties can only be quantified with a joint Bayesian analysis including both interactions.

\begin{figure*}[hbt!]
	\centering
		{\includegraphics[width=\linewidth]{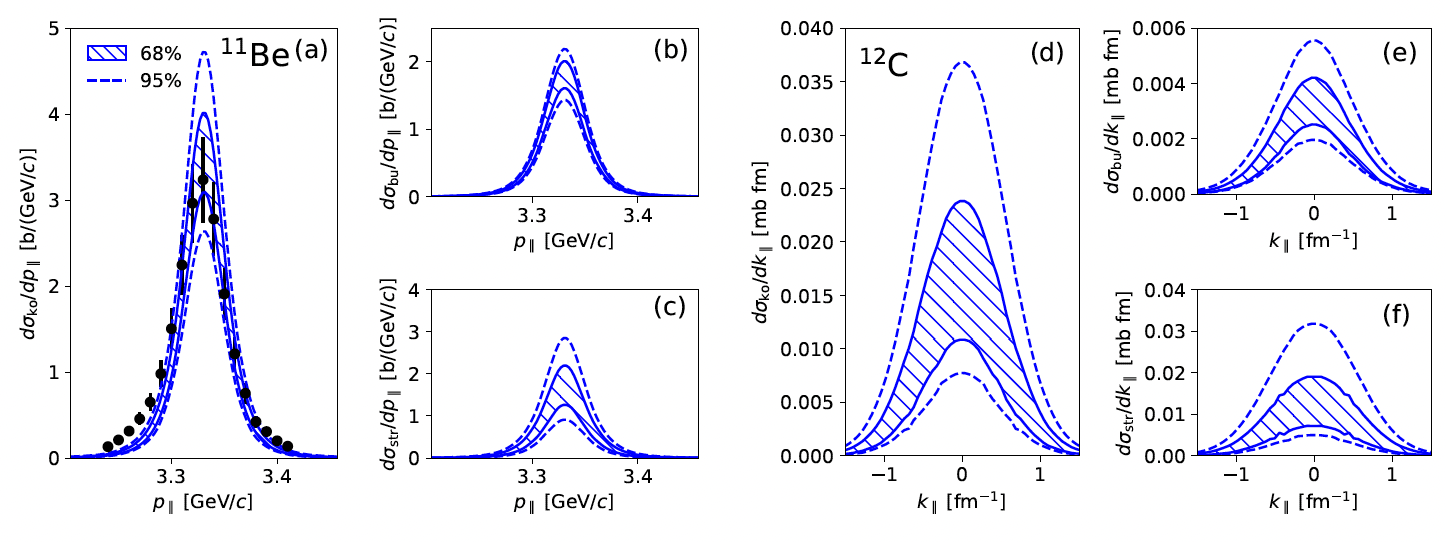}}	
	\caption{{One-neutron knockout cross sections (a) [(d)] $\sigma_{ko}$ with their contributions (b) [(e)] $\sigma_{bu}$ and (c) [(f)] $\sigma_{str}$ for a $^{11}\rm Be$ ($^{12}$C) projectile off a $^9$Be target at 60$A$~MeV. }The black points are experimental data~\cite{PhysRevLett.84.35,PhysRevC.66.024607} and the theoretical bands represent the 68\% (hatched area) and 95\% (dashed lines) uncertainty intervals constructed from the joint Bayesian analysis of the optical potential parameters for the $n$-$^9$Be and $c$-$^9$Be interactions. }\label{Fig4}
\end{figure*}

\subsection{Joint Bayesian analysis of the optical potential parameters of the $n$-$^9$Be and $c$-$^9$Be interactions}

In this section, we consider the uncertainties associated with both the $n$-$^9$Be and $c$-$^9$Be interactions to determine the  theoretical uncertainties on  knockout observables. The parametric uncertainties discussed in Sec.~\ref{Sec3A} and Appendix~\ref{AppB}, determined from fitting the elastic scattering data up to $60^\circ$ for the   $n$-$^9$Be potential  and to $11^\circ$ for the $c$-$^9$Be interaction, were propagated through the eikonal model to knockout momentum distributions and total cross sections. We performed this study by first propagating the uncertainties from only the $n$-$^{9}$Be interaction  (N) and the uncertainties from only the $c$-$^{9}$Be interaction  (C). We then included both simultaneously (NC). As before, we consider the one-nucleon knockout of $^{11}$Be and $^{12}$C on a $^9$Be target at 60$A$~MeV. The $^{11}$Be case has been previously measured~\cite{PhysRevC.66.024607,PhysRevLett.84.35}, but data for $^{12}$C knockout at this energy are not available. We do not consider the uncertainties due to the neutron-core interaction in the projectile, which can be uniquely constrained by structure observables: the ANC for halo nuclei~\cite{PhysRevC.104.024616} and the rms radius for more bound systems~\cite{HCinprep,PhysRevC.77.044306}.

Results for the integrated cross sections for breakup $\sigma_{bu}$, stripping $\sigma_{str}$ and knockout $\sigma_{ko}$ are displayed in Table \ref{t2}, for the $^{11}$Be and $^{12}$C cases. The first row are the results including only the uncertainties from the $n$-$^{9}$Be interaction, the second row are the results with the uncertainties from the $c$-$^{9}$Be interaction and the third row corresponds to the cross sections including both uncertainties. 
Fig. \ref{Fig4}(a) [Fig. \ref{Fig4}(d)] contains the theoretical predictions for the total parallel-momentum distributions $\sigma_{ko}$ for $^{10}$Be ($^{11}$C) detected following the knockout reaction. In Fig. \ref{Fig4}(b) [Fig. \ref{Fig4}(e)] we show the parallel-momentum distributions for breakup $\sigma_{bu}$ (stripping $\sigma_{str}$) for the $^{11}$Be case. Similarly, in Fig. \ref{Fig4}(c) [Fig. \ref{Fig4}(f)] we plot the parallel-momentum distributions for breakup (stripping) for the $^{12}$C case. The hatched bands correspond to the $68$\% uncertainty intervals while the areas delimited by the dashed lines correspond to the $95$\% uncertainty intervals. Finally, we also show the experimental data for the $^{11}$Be case \footnote{The experimental data are extracted from Figs. 11 and 12 of Ref.~\cite{PhysRevC.66.024607}. The error bars correspond to the 15\% experimental uncertainties reported in Ref. \cite{PhysRevLett.84.35}. For a proper comparison between theory and experiment, we have adjusted the position of the center of the parallel-momentum distribution to the data.}.

It is important to emphasize the good agreement with the $^{11}$Be knockout data, both for the integrated cross section and for the momentum distribution, providing confidence that the reaction model used is realistic. Indeed, the eikonal model has been shown to perform well in this case~\cite{PhysRevC.104.024616}. The underestimation of the cross section for low momentum was discussed in Ref.~\cite{PhysRevC.104.024616} and is understood to be caused by dynamical effects not included in the eikonal approximation.

We now turn to uncertainty quantification, the main goal of the current study. We first focus on the $^{11}$Be case.
When inspecting the integrated cross section in Table~\ref{t2}, it is evident that the $n$-$^9$Be interaction dominates the theoretical uncertainties. Also, the uncertainties coming from the two optical potentials, $n$-$^9$Be and $^{10}$Be-$^9$Be, do not add up in quadrature. These two features can be understood due to the strong sensitivity of the knockout cross section to the $n$-$^9$Be absorption and the mechanisms at play during the reaction, as discussed in Sec.~\ref{Sec3A}. Finally, the breakup and stripping components are about the same magnitude but the relative uncertainties on the stripping component are significantly larger than the uncertainties on the breakup component, as seen when analysing the integrated cross sections or by comparing Figs. \ref{Fig4}(b) and (c).
The resulting $1\sigma$ theoretical uncertainties on the predicted knockout cross section of $\approx 15$\% is comparable to the corresponding experimental error bar. 

For the deeply-bound case shown in Fig. \ref{Fig4}(d,e,f), the knockout cross section is dominated by the stripping contribution and its uncertainty. For this case, both $n$-$^9$Be and $^{11}$C-$^9$Be interactions produce large uncertainties in the integrated knockout cross sections (see Table~\ref{t2}). 
One can note  the large differences between the stripping contributions obtained  in the N and C calculations for $^{12}$C. The reduction of these cross sections  in the N case can be explained by the small  imaginary volume radius $r_W$ of the posterior distribution for the  $n$-$^9$Be interaction compared to the $r_W$ of the potential generating the mock data. In the N case, the neutron is therefore less likely to be absorbed than in the C case, and  the stripping contribution is reduced. In the NC case, the stripping cross section is not only influenced by the posterior of the $n$-$^9$Be interaction but also  by the  $r_W$ of the posterior distribution for the  $^{11}$C-$^9$Be interaction, which reflects the probability that the core  survives the collision. Compared to the $r_W$ of the potential generating the  $^{11}$C-$^9$Be mock data, the posterior distributions contain smaller $r_W$. The core  is therefore more likely to survive the collision in the NC case than in the N case, enhancing the stripping cross sections and compensating the reduction caused by the smaller   $r_W$ of the  $n$-$^9$Be interaction.
 These differences between the N, C and NC cases  illustrate the fact that there is no single solution to the inverse scattering problem, i.e., there exists an infinite set of potentials exhibiting different features, such as their geometry and  imaginary strength, reproducing the same scattering data. 
Unfortunately, there are no exclusive knockout data populating the $3/2^-$ ground state of $^{11}\rm C$ at 60$A$~MeV, to compare with our predictions. However, one can compare the theoretical $1\sigma$ uncertainties we obtained due to the optical potentials (about 40-50~\%), to the experimental ones for typical knockout measurements (about 10-15\%)~\cite{PhysRevC.63.024613,Bazinprivate}. These results suggest that the theoretical uncertainties for the tightly bound cases will dominate the overall errors bars.

By comparing Fig. \ref{Fig4}(a) and (d), one may conclude that the relative uncertainties on the knockout cross section will be larger for the knockout of tightly bound neutrons. This result is intuitive given that, for a tightly bound nucleon, there is an increased relevance of small impact parameters in the knockout cross section, which should increase the sensitivity to the details of the optical potentials.  {Additional tests demonstrated that we obtain similar uncertainties for  the one-neutron knockout on $^{10}$C, suggesting that these uncertainties are about 40\% not only on $^{12}$C but also  for  more asymmetric deeply-bound  nuclei. }

The theoretical uncertainties presented here are a lower bound, since there are additional effects not included in our Bayesian analysis. First, the uncertainties introduced by the eikonal description of the reaction, and in particular by the adiabatic and core spectator approximations, are expected to be significant~\cite{PhysRevC.83.011601,PhysRevLett.108.252501}. 
Secondly, any uncertainties associated with the overlap function between initial and final state have not been included. However, this second effect is expected to be small for the two systems considered, as the overlap functions can be well constrained by ANCs and rms radii~\cite{PhysRevC.104.024616,HCinprep}.

\begin{table*}

	\begin{tabular}{cc|ccc} 
	\hline \hline&&$\sigma_{bu}$&$\sigma_{str}$&$\sigma_{ko}$ \\
	&	&[mb]&[mb]&[mb] 	\\\hline 
	\multirow{4}{*}{$^{11}$Be+$^{9}$Be}&N& $87-106\, (79-118)$ & $72-123 \,(53-164)$ &
	$173-222\, (148-261)$ \\
	&C& $98-{110 \,(94-118)}$ &$69-79 \,(65-83)$&$168-189 \,(159-199)$\\
	&NC &$86-107 \,(76-117)$&$68-120\, (49-158)$&$164-217 \,(140-260)$
	\\
	&M&$60-72.7\, (59-89)$&$68-108 \,(52-134)$&$127-142 \,(119-176)$\\
	&Exp.&&&203 $\pm$ 31\\\hline
	\multirow{3}{*}{$^{12}$C+$^9$Be}&N&$3.2-4.4\, (2.7-5.6)$&$1.7-8.1 \,(0.7-15.5)$&$5.5-12.2 \,(4.9-20)$\\
	&C&$3.4-4.9\,(2.9-6.0)$ &$9.8-16.0 \, (7.1-19.0)$ &$13.2-20.9\, (10.2-25.1)$ 
	\\
	&NC &$2.9-5.1\, (2.5-7.2)$&$10.3-27.3 \,(6.1-44.1)$&$13.7-32.0\, (9.3-50.7)$ \\
	&M&$1.8-6.2 \,(1.8-7.2) $&$10.5-32.0\, (8.3-52.9)  $&$11.7-36.4\, (9.9-59.7)$\color{black} \\
	\hline\hline
\end{tabular}	

	\caption{Integrated one-neutron knockout cross sections $\sigma_{ko}$ along with their diffractive-breakup $\sigma_{bu}$ and stripping $\sigma_{str}$ contributions for $^{11}\rm Be$ and {$^{12}$C off a $^9$Be target at 60$A$~MeV}. The first (second) interval  correspond respectively to the $1\sigma$ ($2\sigma$) uncertainties obtained from a Bayesian analysis of the $n$-$^9$Be (N) and the $c$-$^9$Be (C) interaction separately,  a joint analyses of both interactions (NC) and microscopic results (M). 	\label{t2}}
\end{table*}

\section{Microscopic approach} \label{Sec4}

\begin{figure*}[hbt!]
	\centering
		{\includegraphics[width=\linewidth]{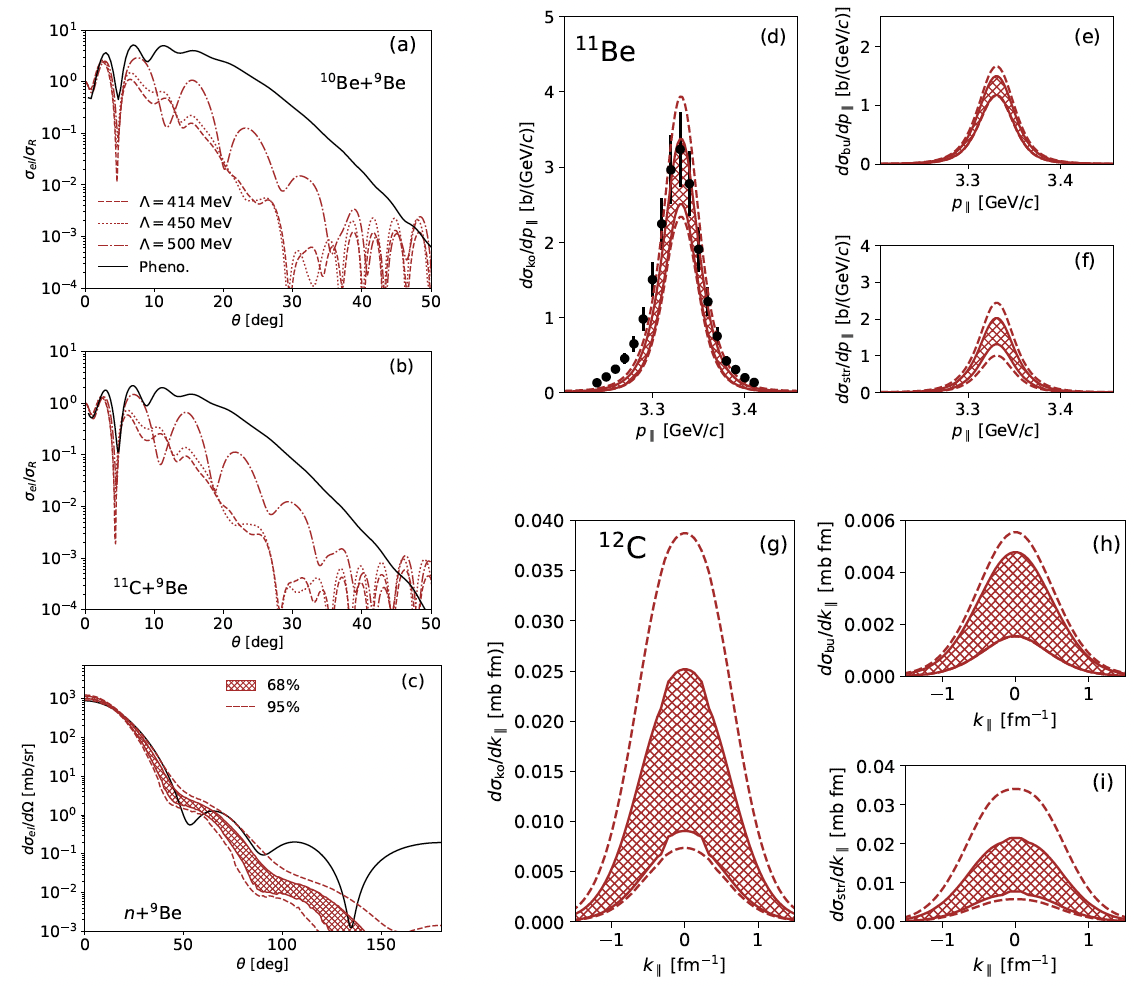}}	
	\caption{Elastic scattering cross sections for $^{10}$Be, $^{11}$C, and $n$ off a $^{9}$Be target at $60A$~MeV are shown in panels (a), (b), and (c) respectively. The black lines are obtained with the phenomenological potentials used to generate the mock data (see Sec.~\ref{Sec3}). For the $^{10}$Be-$^9$Be and $^{11}$C-$^9$Be cross sections, the microscopic predictions for different values of the chiral EFT cutoff are shown in brown. For the $n$-$^{9}$Be cross section in (c), results of 1600 samples of the WLH global optical potential without a spin-orbit term are shown in brown for both 68\% and 95\% uncertainty intervals. {The one-neutron knockout cross section for $^{11}$Be ($^{12}$C) projectiles on a $^{9}$Be target are shown in (d) [(g)] with contributions from breakup (e) [(h)] and stripping (f)~[(i)]. The microscopic knockout cross sections produced by the potentials benchmarked in (a,b,c) are shown in brown and experimental data in black in panel (d). }}\label{FigMicro}
\end{figure*}

In this section, we consider the uncertainties associated with microscopic optical potentials that we propagate to both elastic-scattering and knockout cross sections. Microscopic approaches offer an alternative way of quantifying uncertainties in optical potentials, namely from the uncertainties of nuclear forces and many-body frameworks.

We utilize $n$-$^9$Be and $c$-$^9$Be interactions constructed from 
microscopic nuclear matter calculations of the nucleon self-energy computed from many-body perturbation theory (MBPT) with nuclear forces derived within chiral-EFT. 
For both interactions, the local density approximation is employed to obtain optical potentials for finite nuclei. In the current microscopic approach, there are several sources of theoretical uncertainties. These include uncertainties in chiral effective field theory, truncations in the many-body perturbation expansion, and uncertainties associated with the local density approximation (LDA). Given that there is no clear path to quantify the uncertainties of the LDA, they are not considered in this work. The uncertainties from chiral EFT and MBPT are more straightforward to quantify. In Refs. \cite{Holt17,Drischler19} it has been shown that MBPT uncertainties are smaller than those from chiral EFT for calculations of the nuclear matter equation of state. For the following analysis, we only consider uncertainties from chiral EFT since they are expected to be the dominant theoretical uncertainty.

The $n$-$^9$Be interaction is modeled by the WLH global optical potential, for which a multivariate parameter distribution was generated using a set of chiral forces \cite{Whitehead21}. In practice, one can build up uncertainty intervals for a specific reaction observable by drawing many samples from this distribution. The $c$-$^9$Be optical potential used in this section \cite{whitehead2022} is constructed in a framework that derives a nucleus-nucleus interaction from the energy densities of the interacting nuclei \cite{Brueckner68}. The interaction term of the energy densities is derived from the same nuclear matter self-energy used in WLH. This allows for a consistent construction of nucleon-nucleus and nucleus-nucleus interactions. The $c$-$^9$Be interaction is calculated using a set of three N$^{3}$LO chiral forces with varied cutoffs in the range of $\Lambda = 414 - 500$ MeV. Contrary to the $n$-$^9$Be potential, the $c$-$^9$Be optical potentials have not been parameterized in a global form and therefore cannot be sampled in the same way. The uncertainty intervals for knockout observables to be presented in this section are constructed from the $c$-$^9$Be interactions using each chiral force and  1600 samples of the $n$-$^9$Be potential. {Note that the repulsive character  of these microscopic potentials at short distances, i.e., below 2~fm, do not contribute to the knockout calculations since the absorption dominates at these distances, and the core does not survive the reaction.  }

In Fig.~\ref{FigMicro}, we first look at how these microscopic interactions compare with phenomenological ones used to generate the mock data (solid black line, see Sec.~\ref{Sec2B}) for the scattering of (a) $^{10}$Be, (b) $^{11}$C and (c) $n$ off a $^9$Be target. For Figs. \ref{FigMicro}(a) and (b), the $^{10}$Be-$^{9}$Be and $^{11}$C-$^9$Be angular distributions are obtained for the three different choices for the chiral force (brown dashed, dotted and dot-dashed lines). 
As expected, all microscopic predictions agree with phenomenological ones for forward angles, where the Coulomb repulsion dominates and details of the optical potential play a minor role. The similarity in the diffraction patterns also suggest that the range of the microscopic optical potential is similar to the phenomenological one. 
At larger angles, the microscopic nucleus-nucleus cross sections are smaller than the phenomenological results, indicating that these microscopic interactions are more absorptive at angles below 50$^\circ$. This  larger absorption strength is also consistent with the microscopic results for $n$-$^9$Be scattering which tends to produce cross section smaller than the  phenomenological one [see Fig.~\ref{FigMicro}(c)]. Above 40$^\circ$, the microscopic nucleus-nucleus cross sections diverge, most likely due to the breaking down of the frozen density approximation used to build the microscopic potential, as discussed in Ref.~\cite{whitehead2022}. 

It must be stressed that neither the phenomenological potential that generates the mock data nor the microscopic potential are guaranteed to provide the correct answer over the whole angular range. Since the core-target interaction was fitted to elastic data for a different system and with experimental data only up to $11^\circ$, we cannot determine which, if any, is correct. However, this fact has no impact on our goal which is not to predict elastic scattering but rather to quantify the uncertainty in knockout observables.

We now use these microscopic interactions and propagate their uncertainties to one-neutron knockout cross sections of $^{11}$Be and $^{12}$C, as well as to their breakup and stripping contributions. 
Results for integrated cross sections are shown in Table \ref{t2} with the label (M), and the parallel-momentum distributions are plotted in Fig.~\ref{FigMicro}. 
For the $^{11}$Be case, both the total knockout cross section and momentum distribution [Fig.~\ref{FigMicro}(d)] fall within the experimental error bars. 
However, there are two noticeable differences when compared to the phenomenological results in Fig.~\ref{Fig2}: all the cross sections are slightly smaller and the relative importance of the stripping contribution is larger. Both of these features can be traced back to the more absorptive character of the microscopic interactions. The stripping process is favored because as the neutron is more easily absorbed by the target and there is a decrease the overall knockout strength as the core is less likely to survive the collision.
Similarly, for the $^{12}$C case [Fig. \ref{Fig2}(g,h,i)] the relative importance of the stripping cross section is larger for the microscopic interactions than the phenomenological ones. However, this causes an opposite effect than for the loosely-bound case: the total knockout cross section obtained using the microscopic potentials is now slightly larger than in the phenomenological case.

 Despite the large disagreement between the elastic-scattering angular distribution obtained with microscopic and phenomenological potentials above $10^\circ$, the knockout cross sections have a comparable magnitude and exhibit similar uncertainties. This can be explained by the compensation of the strong absorptive character of the microscopic $n$-$^9$Be and $c$-$^9$Be interactions, suggesting that one may achieve reasonable predictions of knockout cross sections as long as both interactions are derived within the same formalism.

\section{Conclusions} \label{Sec5}

Given the importance of knockout reactions in exploring the structure of rare isotopes at the limits of stability, it is essential to understand the uncertainties in the theoretical models used to interpret those data. This study performs a Bayesian analysis on the reaction model, quantifying parametric uncertainties on the optical potentials, to obtain uncertainty intervals for knockout observables, both momentum distributions and integrated cross sections. We first perform a calibration of the optical model parameters using  mock elastic scattering data as the constraint. Then we propagate the resulting posterior distributions through the knockout reaction framework. We consider the knockout of a halo nucleus $^{11}$Be and that of a tightly bound system $^{12}$C on $^9$Be at 60$A$~MeV to cover the range of possible dynamics. We inspect the two separate components contributing to knockout, namely the breakup component (more sensitive to larger impact parameters) and the stripping component (sensitive to surface effects).

There were two aspects we studied in detail, in preparation for the full analysis: first the relevance of the imaginary surface term relative to the imaginary volume term in the optical potentials, and second the relevance of backward angles in the elastic scattering data used to constrain the interaction.
We find that even at these intermediate energies, it is mostly the surface term of the optical potential that is constrained by elastic scattering data. Having an unconstrained imaginary volume term in the effective interactions can introduce large uncertainties in the knockout predictions, particularly in the stripping component which is sensitive to the imaginary strength. We resolve this ambiguity by following Koning and Delaroche; we take  the geometry of the imaginary volume term to be the same as that of the real volume term. However, it is clear that this uncertainty could be greatly reduce by using {more non-elastic reaction data that  constrain the  imaginary part of these potentials.}
Concerning the importance of backward angles in the elastic angular distributions used to calibrate the optical potentials, while this has a pronounced effect in the uncertainty bands obtained for elastic scattering, their impact on the knockout observables is not convincing. The influence it has on the breakup component is insignificant, and the effect on the stripping component is entangled and depends simultaneously on the overlap function of the projectile and both the $n$-$^9$Be and $c$-$^9$Be optical potentials.

We next perform the full analysis, quantifying both the $n$-$^9$Be and $c$-$^9$Be parametric optical potential uncertainties. These two contributions do not add up coherently and must be evaluated simultaneously. We show that for the one-neutron knockout off a loosely-bound projectile, the parametric $1\sigma$ uncertainty is comparable to the experimental one, i.e. about $15\%$, while for a more deeply-bound projectile, these parametric errors dominate and are about 40\%.

In addition, we also include a microscopic study, in which the optical potentials are obtained from many-body calculations with chiral force. We propagate the uncertainties from chiral EFT directly to the knockout observables, without any calibration to elastic scattering data. The microscopic optical potentials predict larger absorption for the fragment-target interactions used, and there are significant differences in the magnitude of the total knockout cross section. {Although the relative importance of breakup and stripping contributions differ when compared to the phenomenological approach, }our results produce a $1\sigma$ uncertainty band for knockout of the same order of magnitude as the phenomenological approach.

In conclusion, we expect that any information on the overlap function, such as the root mean square radius or the spectroscopic factor, inferred from the ratio of experimental to theoretical knockout cross sections, will carry an uncertainty of at least $\approx \sqrt{0.1^2+0.4^2}\times 100\approx40\%$ (assuming the experimental $10\%$ and theoretical $40\%$ uncertainties are uncorrelated).
For loosely-bound nuclei, if no accurate theoretical prediction exists for the ANC, one can extract it from the ratio of the experimental and theoretical knockout integrated cross sections, when the process is peripheral~\cite{PhysRevC.104.024616,PhysRevC.100.054607}.
The uncertainties on the extracted ANC will derive mainly from the experimental errors and the theoretical uncertainties coming from the optical parameters. 
In the $^{11}$Be example discussed here, the theoretical $1\sigma$ errors are about 15\% and the experimental ones are about 15\%. Thus the $1\sigma$ uncertainties on the ANC$^2$ will therefore be about $\sqrt{0.15^2 +0.15^2}\times100\approx20\%$. 

Furthermore, additional model uncertainties such as the adiabatic approximation and core spectator assumptions may be less accurate, especially for the removal of deeply-bound nucleons~\cite{PhysRevC.83.011601,PhysRevLett.108.252501}. Recent efforts have been devoted to develop new models for knockout relaxing these approximations~\cite{GFK,GOMEZRAMOS2022137252}. Once these new models become available, it will be useful to apply Bayesian tools to perform model comparison between the two approaches.

\begin{acknowledgements}
C.~H. acknowledges the support of the U.S. Department of Energy, Office of Science, Office of Nuclear Physics, under the FRIB Theory Alliance award no. DE-SC0013617 and under Work Proposal no. SCW0498. A.~E.~L. acknowledges the support of the Laboratory Directed Research and Development program of Los Alamos National Laboratory. This work was performed under the auspices of the U.S. Department of Energy by Lawrence Livermore National Laboratory under Contract No. DE-AC52-07NA27344 and by Los Alamos National Laboratory under Contract 89233218CNA000001.
This work was supported by the U.S. Department of Energy grant DE-SC0021422. This work relied on iCER and the High Performance Computing Center at Michigan State University for computational resources.

\end{acknowledgements}

\appendix 
\section{Importance of the surface term}\label{AppA}

\begin{figure*}[hbt!]
	\centering
		{\includegraphics[width=\linewidth]{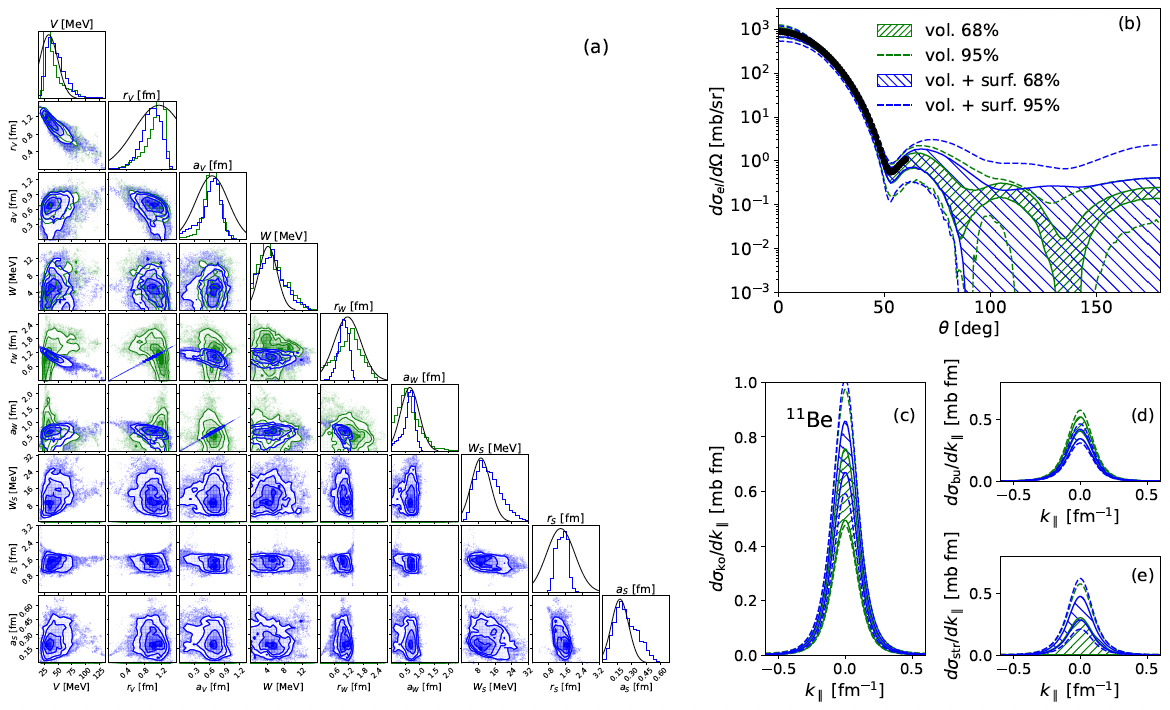}}	
	\caption{Comparison of two Bayesian analyses of the optical potential parameters of the $n$-$^9$Be interaction: (green) volume real and imaginary parameters included in the optimization and (blue) the real volume, imaginary surface, and imaginary volume parameters included in the optimization while imposing the constraint $r_W=r_V$ and $a_W=a_V$. (a) Corner plots; Corresponding 68\% (hatched area) and 95\% (dashed lines) uncertainty bands for (b) elastic-scattering cross section of $n$ off $^{9}\rm Be$ and (c)~one-neutron knockout cross section $\sigma_{ko}=\sigma_{bu}+\sigma_{str}$ of $^{11}\rm Be$ off a $^9$Be target at $60A$~MeV, along with its (d) diffractive-breakup $\sigma_{bu}$ and (e) stripping $\sigma_{str}$ contributions.} \label{Fig1}
\end{figure*}

Here, we investigate the importance of the surface term in the optical potential and how well it can be constrained with elastic scattering data before the parameters are propagated to knockout observables. We first consider the $n$-$^{9}$Be interaction, using two parametrizations for the optical potential. In one, we set the surface depth to zero and only vary the six parameters of the volume term, performing the Bayesian analysis by fitting to elastic scattering data with angles up to 60$^\circ$. The parameter posterior distributions are shown as the green distributions and contours in Fig. \ref{Fig1}(a). Although the posterior distributions for the real depth, radius, and diffuseness are well constrained, the parameters in the imaginary volume term have very wide distributions--sometimes even wider than the prior and beyond the limits of what we typically consider reasonable for optical potentials. In particular, the geometry of the imaginary volume term ($r_W$, $a_W$) is not well constrained, leading to a broad distribution in the depth ($W$).

In panel (b), we show the 68\% (hatched) and 95\% (dashed) uncertainty bands for the $n$-$^{9}$Be elastic-scattering cross section. Even though the imaginary volume terms are fairly unconstrained, the uncertainty intervals are constrained within the angular range covered by the data, and the uncertainties are reasonable at backward angles, particularly when considering the 1$\sigma$ intervals. However, when these potentials are propagated to the knockout observables, as shown for a $^{11}$Be projectile in Fig. \ref{Fig1}(c), the constraint is much poorer, particularly for the stripping contribution [Fig. \ref{Fig1}(e)]. This feature is understandable as the imaginary strength of the interaction quantifies the probability of the neutron to be absorbed by the target. Neither adding more elastic scattering data at backwards angles nor including the reaction cross section data in the fit improves the constraint on the imaginary part of this potential. This indicates that at these high energies (above $60A$ MeV), the imaginary volume geometry is not well constrained by elastic-scattering data. This conclusion  also holds for all of the core-target interactions studied in this work.

Instead, we include both volume and surface imaginary terms in the optimization. It was seen in~\cite{PhysRevC.105.054611} that elastic scattering alone was not enough to constrain the three terms in the optical potential Eq.~\eqref{Eq1},  so following that prescription, we fix the geometry of the imaginary volume term to be the same as that of the real volume term, that is $r_W=r_V$ and $a_W=a_V$ (consistently with Ref.~\cite{Koning2003}). The parameter posterior distributions and contour plots are again shown in Fig.~\ref{Fig1}(a), in blue. We see similar posterior distributions for the real volume potential parameters and, as expected, the geometry of the imaginary volume term is much more strongly constrained. However, while the imaginary surface radius has a narrow posterior distribution, the width of the diffuseness and depth posteriors of this piece of the potential are wider than the prior distributions. These wide distributions also lead to very wide uncertainty intervals on the elastic scattering cross section, shown in Fig. \ref{Fig1}(b), blue hatched and dashed regions. 

The tighter constraints on the imaginary volume and surface radii provided by the elastic scattering data reduce the uncertainties on the stripping cross section [Fig. \ref{Fig1}(e)]; uncertainties of this cross sections are now comparable to the ones on the diffractive breakup contribution [Fig. \ref{Fig1}(d)]. This indicates that knockout observables could provide a tighter constraint on the optical potential than only elastic scattering data. Interestingly, the widths of the uncertainty intervals on the breakup contribution are similar between the two calculations (no surface term and fixed volume geometry) as seen in Fig. \ref{Fig1}(d), but this feature is not surprising as the breakup contribution depends on the probability for the neutron to be scattered by the target, which is well constrained by both interactions by fitting to the elastic scattering data. We do note that $\sigma_{bu}$ is smaller when the volume and surface imaginary potentials are included in the optimization, due to a larger absorption from both imaginary terms.

This conclusion holds also for the other system studied in this work. For the tightly bound $^{12}$C case, we observe that the geometry of the imaginary volume term of the $c$-$^9$Be interaction is also poorly constrained by elastic-scattering data and adding reaction cross section data (not shown here) did not improve the constraint. Nevertheless, since the core survives the collision in the knockout reaction, the effect of the unrealistic geometry of the imaginary part of core-target interaction is less dramatic on knockout cross sections. However, for consistency, we keep the geometry constraint $r_W=r_V$ and $a_W=a_V$ for all reactions studied in this work~{\footnote{We have verified that using a geometry similar to the one of the mock potential, i.e.,  $1.04 \times r_W=r_V$ and $a_W=a_V$ for the $c$-$^9$Be interaction, does not significantly influence  our results.}}.

\section{Influence of the backwards scattering angles for the $c$-$^9$Be interaction} \label{AppB}

In this Appendix, we extend the analysis presented in Sec.~\ref{Sec3A} and we study the effect of the backwards scattering angles in the fit of the $c$-$^9$Be interaction on elastic-scattering and knockout observables.

\begin{figure*}[hbt!]
	\centering
		{\includegraphics[width=\linewidth]{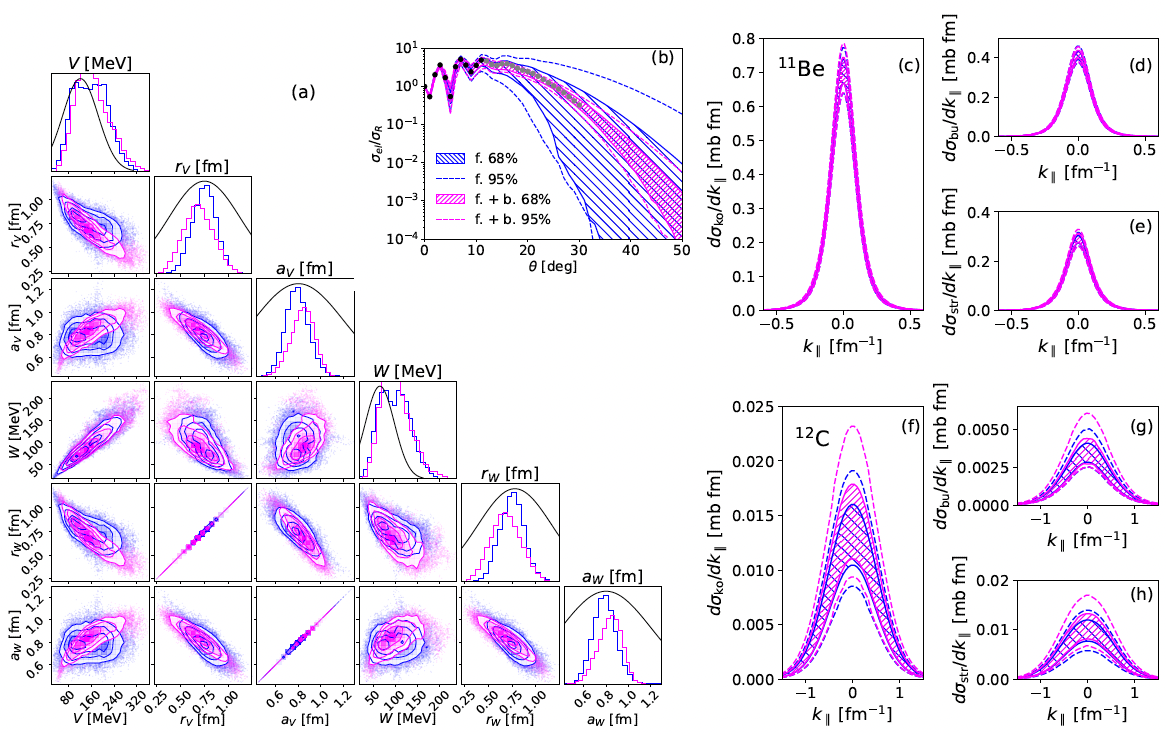}}	
	\caption{Influence of the experimental angular range for the $c$-$^9\rm Be$ interaction on observables: the optical potential parameters are constrained with data $\in [0,11^\circ]$ (f., in blue) and $\in [0,30^\circ]$ (f. + b., in magenta). (a) Corner plots for the $^{10}\rm Be$-$^{9}\rm Be$ optical potential parameters; (b) Elastic-scattering cross sections (normalized to Rutherford) of $^{10}\rm Be$ off $^{9}\rm Be$ as a function of the scattering angle $\theta$ with mock data $\in [0,11^\circ]$ (black points) and $\in [11^\circ,30^\circ]$ (gray points); One-neutron knockout cross sections, (c) $\sigma_{ko}$, and their contributions, (d) $\sigma_{bu}$ and (e) $\sigma_{str}$, of $^{11}\rm Be$ and {(f,g,h) $^{12}$C off a $^9$Be target at $60A$~MeV.} }\label{Fig3}
\end{figure*}

In Fig. \ref{Fig3}, we compare two Bayesian analyses of the $c$-$^9$Be interaction constrained with mock elastic-scattering data at angles $\in [0,11^\circ]$ (in blue) and $\in [0,30^\circ] $ (in magenta). Similar to the analysis in Sec.~\ref{Sec3A}, additional data at backwards scattering angles reduces the 95\% (dashed lines) and 68\% (hatched) uncertainty intervals on the elastic-scattering cross section in this angular range [see Fig. \ref{Fig3}(b)]. However, in this case, the posterior widths stay similar and only the diffuseness and depth parameter correlations are impacted [see Fig. \ref{Fig3}(a)]. We observe similar features for the $^{11}\rm C$-$^9$Be interaction.

On one hand, for the one-neutron knockout of $^{11}$Be [Fig. \ref{Fig3}(c,d,e)], including data at backwards angles for the $^{10}\rm Be$-$^9$Be interaction does not affect the uncertainty intervals for both contributions [Fig. \ref{Fig3}(d,e)] to the total knockout cross section [Fig. \ref{Fig3}(c)]. This can be easily understood by the fact that, as the $^{10}$Be core needs to survive the collision, only contributions for large impact parameters, i.e. forward scattering angles, play an important role for this process. 
 On the other hand, for the removal of a more deeply-bound neutron [Fig. \ref{Fig3}(f,g,h)], including data at larger angles in the fit of the $^{11}\rm C$-$^9$Be interaction increases the uncertainty bands of both the diffractive-breakup [Fig. \ref{Fig3}(g)] and stripping cross sections [Fig. \ref{Fig3}(h)]. As explained in Sec.~\ref{Sec3A}, these contrasting effects for the removal of loosely- and more deeply-bound neutron can be understood by the different mechanisms at play during the reaction. Indeed, for the deeply-bound case, the stripping process is influenced mainly by small impact parameters of both the neutron and the core, which correspond to backwards core-target scattering angles. This analysis confirms the conclusions drawn in Sec.~\ref{Sec3A}, the uncertainties due to both optical potentials do not simply sum up, they can therefore only be quantified with a joint Bayesian analysis involving both interactions. \\

\bibliographystyle{apsrev}
\bibliography{PRCKOreactions}

\end{document}